\documentclass[12pt]{iopart}
\usepackage{iopams}  
\usepackage{bm}
\usepackage{bbm}
\usepackage{graphics}
\usepackage{cite,hyperref}

\begin{document}
	
	\title[\small{Necklaces of $\mathcal{PT}$-symmetric dimers}]{Necklaces of $\mathcal{PT}$-symmetric dimers}
	
	\author{D. J. Nodal Stevens}
	\address{Tecnologico de Monterrey, Escuela de Ingenier\'ia y Ciencias, Ave. Eugenio Garza Sada 2501, Monterrey, N.L., M\'exico, 64849.}

	\author{Benjam\'in Jaramillo \'Avila}
	\address{CONACYT--Instituto Nacional de Astrof\'isica, \'Optica y Electr\'onica, Calle Luis Enrique Erro No. 1, Sta. Ma. Tonantzintla, Pue. CP 72840, M\'exico}
	
	\author{B. M. Rodr\'iguez-Lara}
	\address{Tecnologico de Monterrey, Escuela de Ingenier\'ia y Ciencias, Ave. Eugenio Garza Sada 2501, Monterrey, N.L., M\'exico, 64849.  \\
		Instituto Nacional de Astrof\'isica, \'Optica y Electr\'onica, Calle Luis Enrique Erro No. 1, Sta. Ma. Tonantzintla, Pue. CP 72840, M\'exico}
	\ead{bmlara@itesm.mx}

	\vspace{10pt}
	
	\begin{abstract}
	We study light propagation through cyclic arrays, composed by copies of a given $\mathcal{PT}$-symmetric dimer, using a group theoretical approach and finite element modeling.
	The theoretical mode-coupling analysis suggest the use of these devices as output port replicators where the dynamics are controlled by the impinging light field.
	This is confirmed to good agreement with finite element propagation in an experimentally feasible necklace of passive $\mathcal{PT}$-symmetric dimers constructed from lossy and lossless waveguides.
		\end{abstract}
	
	%
	%
	%
	\maketitle
	%
	%
	
	\section{Introduction}

	Photonic device designers have been toying with the idea of using gain and loss in linear \cite{Somekh1973p46} and non-linear \cite{Chen1992p239} optical systems in order to produce uni-directional couplers or low intensity switches, in that order, for a long time.
	The advent of $\mathcal{PT}$-symmetry in quantum mechanics \cite{Bender1998p5243,Bender2005p277}, the idea that non-Hermitian Hamiltonians invariant under space-time reflection might possess a real spectrum, brought a new structure to these approaches.
	Two seminal ideas made use of this quantum mechanical tool to describe a planar slab waveguide \cite{Ruschhaupt2005p171} and coupled optical structures with symmetric gain and losses \cite{ElGanainy2007p2632}.
	The latter became the true detonator of $\mathcal{PT}$-symmetry research in photonic devices and a plethora of optical proposals followed \cite{HuertaMorales2016p83}. 

\begin{figure}
		\centering
		\includegraphics{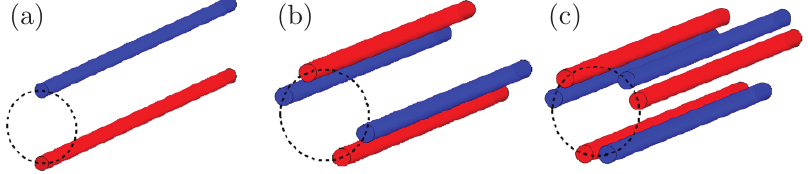}	
		\caption{Schematics for necklaces of $\mathcal{PT}$-symmetric dimers for repetition factors, (a) $N=1$, (b) $N=2$, (c) $N=3$ with intra-dimer couplings smaller than inter-dimer couplings, $g_{d} < g_{N} $.} \label{fig:Fig1}
	\end{figure}

	The quintessential $\mathcal{PT}$-symmetric optical device is the balanced gain and loss dimer, Fig.\ref{fig:Fig1}(a), that has been experimentally realized in diverse optical platforms \cite{Guo2009p093902,Ruter2010p192,Ornigotti2014p065501,Peng2014p394,Hodaei2014p975}. 
	It can be understood as a finite, non-unitary realization of the Lorentz group \cite{RodriguezLara2015p5682} allowing for three types of propagation dynamics: periodic oscillation with amplification, linear, and exponential field amplitude amplification.
	The first case corresponds to the $\mathcal{PT}$-symmetric regime, where eigenvalues are real. 
	The second case is the Kato exceptional point, where the eigenvalues are degenerate and equal to zero. 
	Finally, the third case displays $\mathcal{PT}$-symmetry breaking, where the eigenvalues are purely imaginary. 
	It is also known that any optical realization of the Lorentz group outside the $\mathcal{PT}$-symmetric regime shows asymptotic behaviour, for renormalized intensities, that depends only on the interplay of gain and coupling parameters for both linear \cite{RodriguezLara2015p5682} and nonlinear systems \cite{HuertaMorales2016p83}.
	
	We are interested in a natural extension of the nonlinear $\mathcal{PT}$-symmetric dimer: its periodic repetition over a circular loop.
	Light propagation through these so-called necklaces where individual waveguides are homogeneously spaced have been shown to (dis)allow a $\mathcal{PT}$-symmetric regime for (even) odd repetition of the dimer \cite{Barashenkov2013p033819}.
	The nonlinear four waveguide array is known to produce asymmetric distribution of optical power \cite{Li2011p066608}, possess continuous families of nonlinear modes \cite{Zezyulin2012p213906}, and is not $\mathcal{PT}$-symmetric in the usual matrix sense \cite{Li2012p444021}. 
	The absence of exceptional points, for homogeneous arrays of dimension four, was discussed using linear arrays of four sub-wavelength waveguides \cite{Liu2016p22711}, while the possibility to restore $\mathcal{PT}$-symmetry through mechanical action has also been showed \cite{Longhi2016p1897}.

	Here, we will construct a slightly more general system where the standard $\mathcal{PT}$-symmetric dimer is repeated $N$-times over a circular loop, with the addition of  constant intra- and inter-dimer coupling lengths that are different from each other, Fig. \ref{fig:Fig1}. 
	Such a model can be realized in laser inscribed arrays of waveguides, circular multi-core fibers, toroidal cavities, or electronic systems.
	In the following sections, we introduce the model and show that it can be composed using the cyclic and Lorentz groups. 
	Then, we diagonalize it in the cyclic group basis to show that its dynamics are those of uncoupled effective dimers with variable couplings. 
	We discuss the eigenvalues of these effective dimers and show that homogeneously distributed necklaces with even number of copies always have at least a pair of imaginary eigenvalues, and show that $\mathcal{PT}$-symmetry can be restored by making the intra- and inter-dimer couplings different.
	Then, we show that the underlying cyclic symmetry suggest the use of these photonic devices as output replicators, where input light is used to select the effective dynamics to replicate in the outputs.
	We provide finite element simulations to justify our analytic approach using data from experimental work on waveguide laser inscription in silicon. 
	We close our article with a brief summary and conclusion.

	\section{Linear model and its underlying symmetry}
	
	It is straightforward to note that the necklace is invariant to rotations of $2 \pi /N$ radians around its center; in other words, it has an underlying $\mathbb{Z}_{N}$ symmetry provided by the cyclic group of dimension $N$. 
	It is well-known that the $\mathcal{PT}$-symmetric dimer posses an underlying $SO(2,1)$ symmetry; it is a finite, non-unitary representation of the Lorentz group in $(2+1)-$dimensions \cite{RodriguezLara2015p5682,HuertaMorales2016p83}.
	We can use these facts to write the mode-coupling equations for the necklace,
	\begin{eqnarray}	
	- i \partial_{z} \vert \mathcal{E}(z) \rangle = \hat{H} \vert \mathcal{E}(z) \rangle,
	\end{eqnarray}
	where we have used the notation $\partial_{z}$ to represent the partial derivative with respect to the propagation distance $z$. 
	Here, the field amplitude vector,
	\begin{eqnarray}
	\vert \mathcal{E}(z) \rangle = \sum_{j=0}^{N-1} \sum_{k=0}^{1} \mathcal{E}_{N j + k}(z) ~ \vert j \rangle_{N} \otimes \vert k \rangle_{2} 
	\end{eqnarray}
	is given in terms of the complex field amplitudes at each waveguide, $\mathcal{E}_{n}(z)$ with $n=0,1,\ldots, 2N-1$, and we have used the underlying $\mathbbm{Z}_{N} \otimes SO(2,1)$ symmetry to cast the $n = N j + k$ vector of the standard $2N$-dimensional basis as  the outer product between the $j$-th vector of the standard basis of dimension $N$ and the $k$-th vector of the standard basis of dimension two, $\vert n \rangle = \vert Nj+k\rangle_{2N}=\vert j \rangle_{N} \otimes \vert k \rangle_{2}$.
	The mode-coupling matrix,
	\begin{eqnarray}
	\hat{H} = \mathbbm{1}_{N} \otimes \hat{H}_{d} + g_{N} \left(  e^{i \phi_{N}}  \hat{C}_{N} \otimes \hat{\sigma}_{+} + e^{-i \phi_{N}} \hat{C}_{N}^{\dagger} \otimes \hat{\sigma}_{-}  \right)
	\end{eqnarray}
	is provided by the Kronecker product of the identity matrix of dimension $N$ with the mode-coupling matrix of the twisted dimer, 
	\begin{eqnarray}
	\hat{H}_{d} = i \gamma \hat{\sigma}_{z} + g_{d} \left( e^{-i \phi_{d}} \hat{\sigma}_{+} + e^{i \phi_{d}} \hat{\sigma}_{-} \right). 
	\end{eqnarray}
	We have used the Pauli matrices $\hat{\sigma}_{j}$ with $j=z,\pm$, the effective gain/loss is accounted for by the real parameter $\gamma$, and the intra-dimer coupling by the real parameters $g_{d}$ and $\phi_{d}$. 
	The inter-dimer coupling is provided by the Kronecker product of the shift matrix,
	\begin{eqnarray}
	\hat{C}_{N} = \sum_{j=0}^{N-2}  \vert j+1 \rangle\langle j \vert  +  \vert 0 \rangle \langle N-1  \vert,
	\end{eqnarray}
	and the raising Pauli matrix, $\hat{\sigma}_{+}$, and their adjoint matrices, where the adjoint or conjugate transpose of the shift matrix is given by $\hat{C}_{N}^{\dagger} = \hat{C}_{N}^{N-1}$.  
	The inter-dimer coupling strength and phase are provided by the real parameters $g_{N}$ and $\phi_{N}$, in that order. 
	Note that we have allowed complex coupling parameters, these can be realized by mechanical twisting in optical fibers or capacitive coupling in electronic systems. 
	Also, we obviated the subscript indicating the dimension of the vector space and will do so hereafter.
	
	\section{$\mathcal{PT}$-symmetry and propagation in the linear dimer}

	It is well known that the Fourier matrix \cite{Sylvester1867p461},
	\begin{eqnarray}
	\hat{F}_{N} = \frac{1}{\sqrt{N}} \sum_{j,k=0}^{N-1} e^{i \frac{2 \pi}{N} j k} \vert j \rangle \langle k \vert,
	\end{eqnarray}
	diagonalizes the cyclic group shift operator into the so-called clock matrix,
	\begin{eqnarray}
	 \hat{\Lambda} = \hat{F}_{N} \hat{C}_{N} \hat{F}^{\dagger}_{N} = \sum_{j=0}^{N-1} e^{i \frac{2 \pi}{N} j} \vert j \rangle \langle j \vert.
	\end{eqnarray}
	Thus, we can suggest a change of vector basis provided by the Fourier matrix basis, 
	\begin{eqnarray}
	\vert \mathcal{E}(z) \rangle = (\hat{F}_{N} \otimes \hat{\mathbbm{1}}_{2})^\dagger \vert \mathcal A(z) \rangle,
	\end{eqnarray}
	and recover effective propagation dynamics,
	\begin{eqnarray}
	-i \partial_{z} \vert \mathcal{A}(z) \rangle = \hat{H}_{D} \vert \mathcal{A}(z) \rangle 
	\end{eqnarray}
	described by a block-diagonal mode-coupling matrix,
	\begin{eqnarray}
	\hat{H}_{D} = \sum_{j=0}^{N} \vert j \rangle \langle j \vert \otimes \hat{H}_{j},
	\end{eqnarray}
	where each diagonal block describes a $\mathcal{PT}$-symmetric dimer whose effective coupling depends on its position in the diagonal, 
	\begin{eqnarray}
	\hat{H}_{j} = i \gamma \hat{\sigma}_{z} +  \left[ \Gamma_{j} \hat{\sigma}_{+}  + \Gamma_{j}^{\ast} \hat{\sigma}_{-}\right], 
	\end{eqnarray}
	with the new effective complex coupling given by the following expression,
	\begin{eqnarray}
	\Gamma_{j} =  g_{d} e^{i \phi_{d}}  + g_{N}  e^{- i \phi_{N}}  e^{i \frac{2 \pi}{N} j}.
	\end{eqnarray}
	Here, it is straightforward to calculate the eigenvalues of these matrices,
	\begin{eqnarray}
	\lambda_{\pm j} &=& \pm \sqrt{ \vert \Gamma_{j} \vert^{2} - \gamma^{2}}, \nonumber \\
	&=& \pm \sqrt{ g_{d}^2 + g_{N}^2 + 2 g_{d} g_{N} \cos \left( \phi_{d} + \phi_{N} - \frac{2 \pi}{N} j   \right)- \gamma^2},
	\end{eqnarray}
	 and realize that different intra- and inter-dimer complex couplings can restore the $\mathcal{PT}$-symmetric phase to systems lacking it, for example homogeneously coupled arrays with even repetition number where $g_{d}=g_{N}=g$, $\phi_{d}= \phi_{N} = 0$, and $N=2 n$ with $n=1,2,3, \ldots$ \cite{Barashenkov2013p033819}.
	
\begin{figure}
		\centering
		\includegraphics{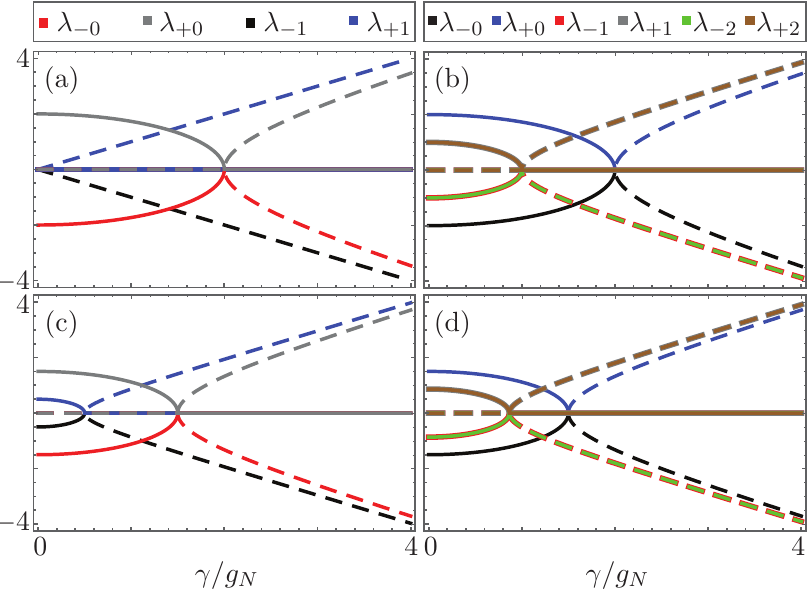}	
		\caption{Real (solid lines) and imaginary (dashed lines) parts of eigenvalues as a function of the gain to inter-dimer coupling ratio, $\gamma / g_{N}$, for arrays of coupled $\mathcal{PT}$-symmetric dimers with homogeneous intra- and inter-dimer couplings, $g_{d} = g_{N}$ with repetition factor (a) $N=2$ and (b) $N=3$, and for inhomogeneous couplings, $g_{d} = g_{N}/2$ with repetition factor (c) $N=2$ and (d) $N=3$. } \label{fig:Fig2}
	\end{figure}

	In homogeneously coupled arrays, the block $\hat{H}_{0}$ is the standard $\mathcal{PT}$-symmetric dimer with Kato exceptional point $\gamma = 2 g$.
	Each block will have its own exceptional point at the following values of the gain to coupling ratio,
	\begin{eqnarray}
	\frac{\gamma}{2 g} = \cos \frac{2 \pi}{N} j, \quad j=0, 1, 2, \ldots \left\lfloor \frac{N}{2} \right\rfloor \left( \left\lfloor \frac{N}{2} \right\rfloor - 1  \right),
	\end{eqnarray}
	for even (odd) dimensions of the cyclic group component; the floor operation $\lfloor \cdot \rfloor$ provides the lowest integer that is less or equal to its argument.
	Then, we can realize that arrays where the dimer is repeated an even number of times always have a pair of complex eigenvalues, $\lambda_{\pm N/2} = \pm i \gamma$, that correspond to the block $\hat{H}_{N/2}$; these arrays do not have a $\mathcal{PT}$-symmetric phase.
	We can also see that there are pairs of diagonal blocks that are the transpose of each other, $\hat{H}_{k} = \hat{H}_{N-k}^{T}$ with $k=1,2,\ldots, \lfloor N/2 \rfloor -1 ~ ( \lfloor N/2 \rfloor)$ for even (odd) dimension.
	This immediately points to the fact that there will be  $\lfloor N/2 \rfloor - 1$  $(\lfloor N/2 \rfloor)$  two-fold degenerate eigenvalues in the arrays where the repetition number $N$ is even (odd).
	Furthermore, the number of imaginary eigenvalues will increase each time the gain to coupling ratio reaches the value of a block exceptional point.
	Figure \ref{fig:Fig2} shows the real to complex eigenvalue transition for arrays of four, Fig. \ref{fig:Fig2}(a), and six, Fig. \ref{fig:Fig2}(b), waveguides.
	The simple change of inter-dimer coupling is enough to induce a $\mathcal{PT}$-symmetric phase in arrays with even repetition rate, but the behaviour of the degeneracies and the systemic increase in the number of complex eigenvalues remains the same.
	Figure \ref{fig:Fig2}(c) and Fig. \ref{fig:Fig2}(d) show the eigenvalues of an array where the intra-dimer couplings are half the value of the inter-dimer couplings, $g_{d} = g_{N}/2$.
	Controlling couplings phases introduces another physical parameter to control the type of eigenvalues in the system.
	
	It is also straightforward to construct a propagator with the block diagonal matrix form, $\hat{H}_{D}$. 
	In the original frame, the propagation of the complex field amplitudes vector,
	\begin{eqnarray}
	\vert \mathcal{E}(z) \rangle = \hat{U}(z)  \vert \mathcal{E}(0) \rangle,
	\end{eqnarray}
	is provided by the propagator matrix, 
	\begin{eqnarray}
	\hat{U}(z) = \sum_{j=0}^{N-1} \hat{F}^{\dagger}_{N} \vert j \rangle \langle j \vert \hat{F}_{N} \otimes e^{ i \hat{H}_{j} z },
	\end{eqnarray}
	in terms of the propagation matrix of each individual effective dimer \cite{HuertaMorales2017p587},
	\begin{eqnarray}
	e^{ i \hat{H}_{j} z } = \hat{\mathbbm{1}}_{2} \cos \Omega_{j} z  + i z \hat{H_{j}} \textrm{sinc}~\Omega_{j} z 
	\end{eqnarray} 
	where we the propagation constant, 
	\begin{eqnarray}
	\Omega_{j} =  \sqrt{ \vert \Gamma_{j} \vert^{2} - \gamma^{2}},
	\end{eqnarray}
	is just the positive eigenvalue of the mode-coupling matrix.
	It takes real values in the $\mathcal{PT}$-symmetric regime, providing a non-unitary oscillator,
	\begin{eqnarray}
	e^{ i \hat{H}_{j} z } = \hat{\mathbbm{1}}_{2} \cos \Omega_{j} z  + \frac{i}{\Omega_{j}} \hat{H_{j}} \sin \Omega_{j} z , \qquad \Omega_{j} \in \mathbb{R}.
	\end{eqnarray} 
	At Kato's exceptional point it has zero value and provides lineal amplification and attenuation of the field amplitudes, 
	\begin{eqnarray}
	e^{ i \hat{H}_{j} z } = \hat{\mathbbm{1}}_{2}  + i z \hat{H_{j}}, \qquad \Omega_{j}= 0.
	\end{eqnarray} 	
	Finally, it takes purely imaginary values in the broken symmetry regime, providing exponential amplification and attenuation of the field amplitudes,
	\begin{eqnarray}
	e^{ i \hat{H}_{j} z } = \hat{\mathbbm{1}}_{2} \cosh \vert\Omega_{j}\vert z  - \frac{1}{\vert \Omega_{j} \vert} \hat{H_{j}} \sinh \vert\Omega_{j}\vert z , \qquad \Omega_{j} \in \mathbb{C}.
	\end{eqnarray}

	\section{Dimer output replication}

	Cyclic symmetry has a curious signature, we can address specific block diagonal elements and produce $N$ copies of its dynamics.
	In our necklace, this is obtained by the following impinging field amplitude vector,
	\begin{eqnarray}
	\vert \mathcal{E}(0) \rangle &=&  \hat{F}_{N}^{\dagger} \vert j \rangle \otimes \vert \psi (0) \rangle, \nonumber \\
	  &=& \frac{1}{\sqrt{N}} \sum_{a}^{N-1} e^{i \frac{2 \pi}{N} a j}  \vert a \rangle \otimes \vert \psi(0) \rangle;
	\end{eqnarray}
	 that is, the outer product between a unit vector, whose components are all the $N$th roots of unity, and some initial state for the effective $j$th block diagonal dimer we want to address.
	 Such an initial state produces replication, up to a phase, of the dynamics provided by the $j$th effective dimer,
	 \begin{eqnarray}
	 \vert \mathcal{E}(z) \rangle &=&  \frac{1}{\sqrt{N}} \sum_{p=0}^{N-1} e^{i \frac{2 \pi}{N} j p} \vert p \rangle \otimes e^{i \hat{H}_{j} z } \vert \psi(0) \rangle,
	 \end{eqnarray}
	 in each and every one of the dimers in the necklace.
	 A simple permutation of input phases allows to switch the output of the array, provided that one doesn't switch to the transposed effective dimer, $\hat{H}_{j} = \hat{H}_{N-j}^{T}$.
	 
	 It is well-known that outside the $\mathcal{PT}$-symmetric regime, the renormalized power at the $k$th waveguide of the $j$th dimer \cite{RodriguezLara2015p5682,HuertaMorales2016p83},
	 \begin{eqnarray}
	 P_{j,k}(z) = \frac{ \vert \langle k \vert   e^{i \hat{H}_{j} z } \vert \psi(0) \vert^2}{\sum_{k=0}^{1} \vert \langle k \vert   e^{i \hat{H}_{j} z } \vert \psi(0) \vert^2}, \quad 
	 \end{eqnarray} 
	 will have an asymptotic behavior, 
	 \begin{eqnarray}
	 \lim_{z \rightarrow \infty} P_{j,0}(z) &=& \frac{\gamma - \vert \Omega_{j} \vert}{2 \gamma}, \nonumber \\
	 \lim_{z \rightarrow \infty} P_{j,1}(z) &=& \frac{\gamma + \vert \Omega_{j} \vert}{2 \gamma}, \label{eq:AsymptPower}
	 \end{eqnarray}
	 dictated by the parameters of the corresponding effective dimer $\hat{H}_{j}$. 
	 Thus, we can engineer photonic devices where all the dimers present the same output, and switch from  oscillatory to asymptotic behaviors by permutation of the input field phases.
	 
\begin{figure}
		\centering
		\includegraphics{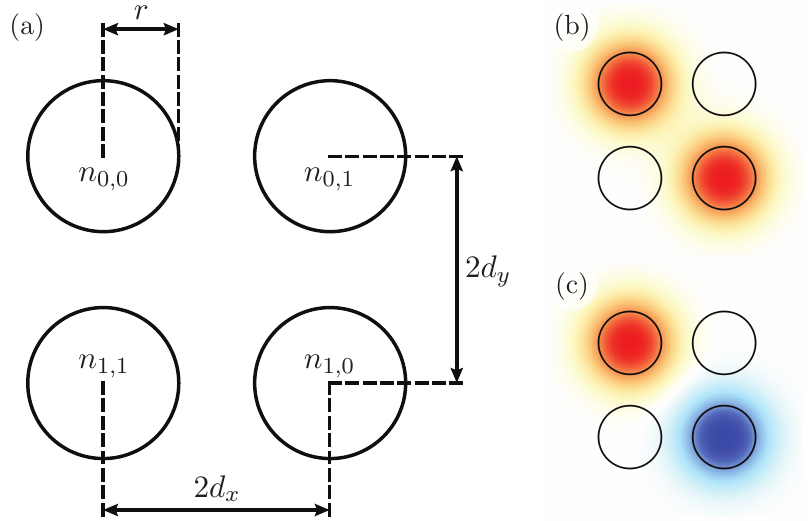}	
		\caption{(a) Waveguide geometry used in the finite element simulation and electric field input for replication of the (b) first and (c) second effective dimers  } \label{fig:Fig3}
	\end{figure}

	 This type of dimer output port replication can be seen in any plausible realization of our model. 
	 For example, let us consider the case of laser inscribed waveguides in silicon, c:Si, for telecommunication wavelength, $\lambda = 1550~ \mathrm{nm}$.
	 Current techniques allow for control of refractive index properties, $\Delta n = n_{\mathrm{core}} - n_{\mathrm{bulk}} \in [1,6] \times 10^{-3}$, as well as losses, $[0.1, 3] \mathrm{dB}/\mathrm{cm}$, with waveguide length of about a centimetre \cite{Nejadmalayeri2005,Chambonneau2016}.
	 We use the commercial finite element software COMSOL to simulate a necklace of two identical passive $\mathcal{PT}$-symmetric dimers realized by four waveguides, radius $r=2.230 ~\mu\mathrm{m}$, in the configuration shown in Fig. \ref{fig:Fig3}(a).
	 The dimers are realized by two lossy waveguides with complex refractive indices $n_{0,0}=n_{1,0}= 3.465 + i ~6.365 \times 10^{-4}$, and two lossless waveguides with real refractive indices $n_{0,1}=n_{1,1}= 3.465$.
	 We consider as input the superposition of two Gaussian beams, each a plane wavefront polarized in the vertical direction with waist radius $\omega_{0} = 3.154~ \mu \mathrm{m}$, impinging at the lossy fiber of each dimer. 
	 The effective $\mathcal{PT}$-symmetric dimer shows an approximated gain/loss $\gamma = 860 ~m^{-1}$.
	 In order to replicate the output of the effective dimer $\hat{H}_{0}$, we use an in-phase superposition of the Gaussian beams, Fig. \ref{fig:Fig3}(b), and a $\pi$-dephased superposition, Fig. \ref{fig:Fig3}(c), to replicate the output of the effective dimer $\hat{H}_{1}$.
	 We retrieve from the simulation the vertical component of the electric field at the propagation axis of each waveguide, and use them to calculate the renormalized power within each dimer,
	 \begin{eqnarray}
	 P_{j,k}(z) = \frac{ \vert E_{y}(  d_{x} (-1)^{j+k}, d_{y} (-1)^j, z) \vert^2}{\vert E_{y}(  d_{x} (-1)^{j}, d_{y} (-1)^j, z) \vert^2 + \vert E_{y}(  d_{x} (-1)^{j+1}, d_{y} (-1)^j, z) \vert^2}.
	 \end{eqnarray} 
	 
\begin{figure}
		\centering
		\includegraphics{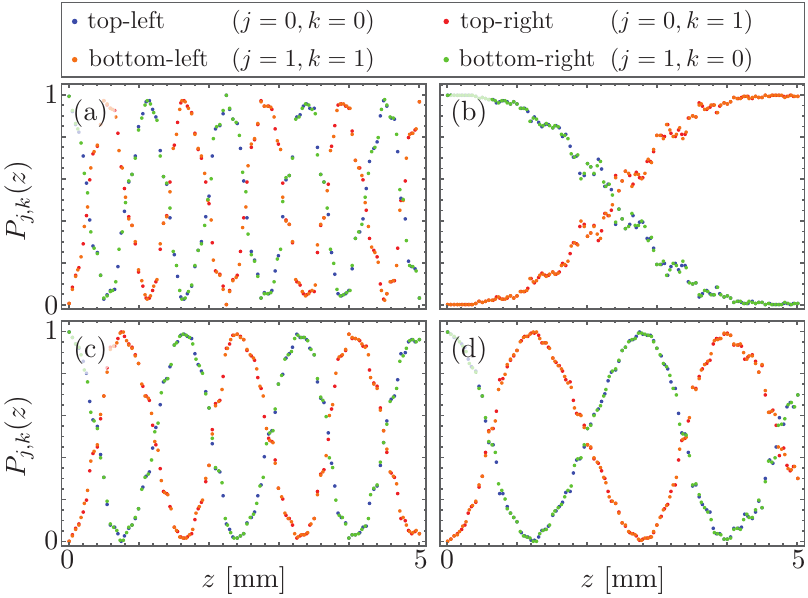}	
		\caption{Renormalized power propagation, $P_{j,k}(z)$, showing output replication for (a-b) homogeneous and (c-d) inhomogeneous couplings. See text for details.   } \label{fig:Fig4}
	\end{figure}

	In the homogeneous coupling configuration, $d_{x}=d_{y}= 2.790 ~\mu \mathrm{m}$, addressing the first effective dimer provides oscillations of the renormalized power with approximated propagation constant $\Omega_{1} \approx 2845.374 ~ \mathrm{m}^{-1}$ as shown in Fig. \ref{fig:Fig4}(a).
	Addressing the second effective dimer provides the switching, \ref{fig:Fig4}(b), predicted from the asymptotic  renormalized power, Eq.(\ref{eq:AsymptPower}).
	In the inhomogeneous coupling configuration, $d_{x}=2.790 ~\mu \mathrm{m}$ and $d_{y} = 3.906 ~\mu \mathrm{m}$, addressing the first and second effective dimers provide renormalized power oscillations with approximated propagation constants $\Omega_{1} \approx 1920 ~\mathrm{m}^{-1}$, Fig. \ref{fig:Fig4}(c), and $\Omega_{2} \approx 1200 ~\mathrm{m}^{-1}$, Fig. \ref{fig:Fig4}(d), in that order, as predicted by coupled-mode theory.
	
	\section{Conclusions}
	
	We have presented a group theoretical analysis for the mode-coupling description of light propagating through so-called necklaces of $\mathcal{PT}$-symmetric dimers.
	We showed that the coupled-mode matrix comprising the field dynamics in these arrays can be written as a composition of elements of the cyclic group and the (2+1)-dimensional Lorentz group.
	Using the fact that the cyclic subcomponent is straightforward to diagonalize, we demonstrated that the necklace effective dynamics are provided by $N$ uncoupled $\mathcal{PT}$-symmetric dimers, where $N$ is the number of dimer copies in the necklace.  
	For even number of copies and equal inter- and intra-dimer couplings, there is always at least one effective dimer in the broken symmetry phase.
	The $\mathcal{PT}$-symmetry of these necklaces can be restored by increasing their radius, making the inter- and intra-dimer couplings different in value.
	In general, for necklaces with three or more dimer copies, there will always be pairs of effective dimers that are the complex conjugated of each other and will show identical light power propagation.
	
	We compared our effective mode-coupling analysis with finite element simulations to good agreement in all cases.
	Our analytic results and computational simulations show that it is possible to use impinging light to switch between the different effective dynamics provided by a necklace of $\mathcal{PT}$-symmetric dimers, and replicate their outputs all over the necklace. 
	Furthermore, our results demonstrate that it is plausible to design devices where oscillations in the $\mathcal{PT}$-symmetric regime can be switched to linear and exponential field amplitude dynamics in Kato exceptional point and broken-symmetry regime, respectively, by changing the relative phases of input fields.
	While our results were simulated using experimental parameters from laser inscribed waveguides, they are valid for multi-core fibers, toroidal cavities, and electronic devices as well.	

	\vspace{6pt} 
	
	\ack{	
		D. J. Nodal Stevens acknowledges financial support from Red de Tecnolog\'ias Cu\'anticas through the program Verano de Investigaci\'on Cient\'ifica and the Photonics and Mathematical Optics group hospitality.
		B. M. Rodr\'iguez-Lara acknowledges support from the Photonics and Mathematical Optics Group at Tecnologico de Monterrey and Consorcio en \'Optica Aplicada through CONACYT FORDECYT $\#$290259 project grant. 
	}

	\section*{References}

%
\end{document}